# Challenges and Opportunities on Using Games to Support IoT Systems Teaching

Bruno P. de Souza, Claudia M. L. Werner
PESC/COPPE
Federal University of Rio de Janeiro - UFRJ
Rio de Janeiro, Brazil
{bpsouza, werner}@cos.ufrj.br

*Abstract*— Context: New systems have emerged within the Industry 4.0 paradigm. These systems incorporate characteristics such as autonomy in decision making and acting in the context of IoT systems, continuous connectivity between devices and applications in cyber-physical systems, omnipresence properties in ubiquitous systems, among others. Thus, the engineering of these systems has changed, drastically affecting the manner of their construction process. In this context, to identify simple and playful alternatives to teach how to build them is a difficult task. Objective: To present how to teach IoT systems using games, to reveal challenges and opportunities obtained through a literature review. Method: A Structured Literature Review (StLR), supported by the Snowballing technique, was executed to find empirical studies related to teaching, games and IoT systems. Results: 12 papers were found about teaching IoT systems using games. As challenges and opportunities, many issues were identified related to IoT systems programming, modularity, hardware constraints, among others. Conclusion: In this work, research challenges and opportunities were found in the context of IoT systems teaching. Due to specific features of these systems, teaching their construction is a difficult activity to carry out.

*Keywords—industry 4.0, internet of things, teaching, literature review*

*Resumo*— Contexto: Novos sistemas têm surgido dentro do paradigma da Indústria 4.0. Esses sistemas são caracterizados por autonomia na tomada de decisão e atuação no contexto de sistemas IoT, conectividade contínua entre dispositivos e aplicações em sistemas ciber-físicos, propriedades de onipresença em sistemas ubíquos. Com isso, a engenharia desses sistemas mudou, afetando drasticamente a forma de seu processo de construção. Neste contexto, identificar alternativas simples e lúdicas para ensinar como construí-los é uma tarefa trabalhosa. Objetivo: Apresentar os desafios e oportunidade em ensinar sistemas IoT usando jogos, obtidos através de uma revisão da literatura. Método: Uma Revisão Estruturada da Literatura (REsL), apoiada pela técnica Snowballing, foi executada para encontrar estudos empíricos relacionados ao ensino, jogos e sistemas IoT. Resultados: 12 artigos foram identificados no ensino de sistemas IoT usando jogos. Foram identificados alguns desafios e oportunidades relacionados à programação de sistemas IoT, modularidade, restrições de hardware, dentre outros. Conclusão: Neste trabalho, os desafios e oportunidades de pesquisa foram encontrados no contexto do ensino de sistemas IoT. Devido às características específicas desses sistemas, o ensino de sua construção se torna uma atividade difícil de realizar.

*Palavras-chave — indústria 4.0, internet das coisas, ensino, revisão da literatura*

## I. Introdução

De acordo com Motta, De Oliveira e Travassos [1], os sistemas IoT são definidos como "um paradigma que promove o entrelaçamento de tecnologias e dispositivos, além de possuírem a capacidade de capturar, trocar e processar dados por meio de uma rede sem fio, tomar decisões e atuar no ambiente, unindo os mundos real e virtual por meio de software."

Sistemas IoT juntamente com sistemas ubíquos e pervasivos, sistemas sensíveis ao contexto, sistemas ciber-físicos têm conduzido a sociedade a quarta revolução industrial, que também é conhecida como Indústria 4.0 [2] [3].

Como consequência da Indústria 4.0, sistemas IoT trazem consigo novos desafios e preocupações no que diz respeito a engenharia do seu ciclo de desenvolvimento (dos requisitos a implantação), na qual decidir qual técnica de elicitação de requisitos utilizar ou o tipo de arquitetura adequada adotar no processo de desenvolvimento torna-se uma tarefa desafiadora [1] [2]. Além disso, tais sistemas também possuem propriedades específicas de hardware e software para serem abordadas, como sua autonomia total ou parcial, grande massa de dados, segurança no acesso a informações críticas, conectividade contínua entre dispositivos, além da interação entre sistemas, usuários e coisas [1] [2].

Zambonelli [4] relata que ainda há pouca evidência disponível na literatura técnica sobre o apoio de tecnologias de software (e.g., técnicas de requisitos, *framework* para codificação, técnicas de testes, dentre outros) para incorporar no ciclo de desenvolvimento dos sistemas IoT. Uma alternativa para este problema é a adoção de tecnologias já consolidadas disponíveis para apoiar a construção de sistemas tradicionais (*web* e móveis). No entanto, talvez, estas tecnologias podem não apoiar completamente a construção dos sistemas IoT, necessitando assim adaptá-las, ou, até mesmo, criar tecnologias de software específica para construção de sistemas IoT.

Devido a essas preocupações advindas da construção dos sistemas IoT, pode-se mencionar a Engenharia de Software (ES), sendo esta uma área da computação que trabalha com o ciclo de desenvolvimento de qualquer tipo de sistema, usando abordagens, técnicas, métodos e ferramentas para isso. A ES também considera aspectos de colaboração entre teoria e a prática para que seu processo construtivo seja realizado de forma completa [5]. Todavia, em sua maioria, educadores enfrentam alguns desafios no ensino de ES. Tais desafios estão associados a conteúdos, quase que totalmente teóricos (sem a parte prática), devido ao tempo de execução de projetos, deficiência na matriz curricular da disciplina, ou até mesmo a falta de conhecimento da prática. Desta forma, encontrar maneiras inovadoras de levar ES prática para dentro de sala de aula seria uma vantagem tanto para os alunos quanto para educadores.

Diante desse contexto, os jogos surgem como uma alternativa de estratégia contemporânea de ensino-aprendizagem que podem resolver parte dos problemas da disciplina de ES, e, consequentemente, da construção de sistemas IoT. Jogos, por sua vez, são definidos como uma atividade recreativa, lúdica e com regras, onde o objetivo principal é o entretenimento do jogador/participante (Dicionário Aurélio).

Assim, uma potencial combinação da área de jogos e sistemas IoT se mostra promissora, devido às especificidades de tais sistemas, bem como o uso lúdico dos jogos para ensino-aprendizagem [18]. Portanto, com o objetivo de ampliar a compreensão do ensino de sistemas IoT para além da utilização de jogos como estratégia de ensino-aprendizagem, este trabalho discute os desafios e oportunidades na utilização de jogos no ensino de sistemas IoT. Assim, foi realizada uma Revisão Estruturada da Literatura (REsL) para alcançar os resultados pretendidos.

Este artigo está organizado da seguinte forma. A Seção II descreve a fundamentação teórica. A Seção III detalha a Revisão da Literatura Estruturada realizada no contexto do uso de jogos para ensinar sistemas IoT. A Seção IV relata os principais resultados encontrados, bem como sua discussão. Em seguida, a Seção V descreve as limitações e ameaças à validade deste estudo. Por fim, a Seção VI conclui apresentando trabalhos futuros e considerações finais.

## II. Fundamentação Teórica e Trabalhos Relacionados

### A. Internet das Coisas (Internet of Things - IoT)

O termo IoT foi inicialmente mencionado por Kevin [20], em que ele se refere a IoT como coisas que são unicamente edificáveis por sensores e que suas representações são realizadas de maneira virtual por uma conexão (internet). Desde então a IoT vem apresentando variáveis definições na literatura técnica nos últimos anos [1] [6] [7]. Os sistemas IoT são objetos interconectados por uma rede que permitem conexão com a internet em qualquer lugar e a qualquer momento. Segundo Atzori *et al.* [6], a ideia básica desse paradigma é a presença de uma ampla variedade de objetos – como sensores, *wearables* atuadores, celulares, entre outros, que, por meio de esquemas de endereçamento exclusivos (*tags*), podem interagir um com o outro.

Motta, Silva e Travassos [7] realizaram uma revisão da literatura com o objetivo de definir sistemas IoT, suas características, domínios e aplicações. Os resultados mostram que a IoT representa um paradigma que permite a configuração entre sistemas de software, hardware (coisas) e permite a conexão à internet a todo instante. Além disso, foi identificado que sistemas IoT possuem quatro comportamentos principais: identificação, sensoriamento, processamento e atuação. Os autores descrevem que é desafiador construir os cenários que um sistema IoT pode possuir, devido aos seus diversos comportamentos dinâmicos e interação entre máquina, coisas, usuários finais e outros sistemas de software.

Embora os sistemas IoT tenham várias definições na literatura técnica, é importante destacar que uma característica comum nessas definições é a presença de coisas conectadas aos sistemas de software pela rede.

### B. Jogos

De acordo com o dicionário Michaelis, jogos são definidos como "Qualquer atividade recreativa que tem por finalidade entreter, divertir ou distrair." Xexéo *et al.* [8] conceitua jogos como atividades voluntárias no âmbito social e cultural, onde é utilizado um "mundo abstrato", levando em consideração efeitos negociados do mundo real, tendo um final imprevisível. Castro, Costa e Werner [9] descrevem que os jogos possuem características interativas, lúdicas e divertidas, o que acaba se tornando um método adequado para fazer com que os alunos absorvam mais o conteúdo de uma forma diferenciada e atual.

De acordo com Adams e Rollings [10], jogos são operações reais, em que os jogadores buscam alcançar os objetivos pré-estabelecidos, seguindo as regras previamente definidas de modo voluntário. De acordo com Kapp [11], jogo pode ser considerado como um sistema em que há 'jogadores (pessoas)' que se envolvem em um determinado desafio, além disso são definidas regras que devem ser seguidas, tendo em resultado final quantificável, e muitas vezes uma reação 'emocional' é ocasionada.

### C. Trabalhos Relacionados

Em [18], Henry et al. propõem um *framework* para projetar jogos sérios inteligentes (ou *smart serious games - SSG*), cujo objetivo é facilitar a projeção e construção de jogos baseados em IoT. Este *framework* é composto pela área de jogos sérios e o tópico emergente da IoT. O SSG é um *framework* composto por cinco camadas: aplicação, processamento de dados, *middleware*, rede e sensoriamento. Os autores validaram o framework através de experimentos com alunos de graduação.

No trabalho de Kassab, Defranco e Laplante [19], os autores realizam uma revisão da literatura para levantar os principais benefícios e desafios em incorporar IoT na educação. Os autores analisam os benefícios e desafios em várias perspectivas. Em relação aos benefícios, os autores mapearam os resultados em um esquema de classificação tridimensional: percepção, princípios de aprendizagem e modo de entrega. Em relação aos desafios, citam alguns já conhecidos como interoperabilidade e big data. Porém, os principais desafios achados foram segurança, escalabilidade e humanização. Kassab, Defranco e Laplante [19] relatam que incorporar IoT na área da educação requer uma série de cuidados a serem considerados.

## III. Método de Pesquisa

Foi realizada uma Revisão Estruturada da Literatura (REsL) para investigar o tema de pesquisa abordado neste estudo. Uma REsL é uma variação das Revisões Sistemáticas da Literatura (RSL). A REsL reutiliza algumas características da RSL, como o protocolo de pesquisa, procedimento de seleção (critérios de inclusão e exclusão), procedimento de extração de informações, dentre outros. Por outro lado, outras características da RSL não foram incorporadas, como a avaliação da qualidade do artigo ou o uso de alguma técnica sofisticada de análise de dados (como a *grounded theory*). Além disso, outra característica da REsL é um menor tempo na execução que uma RSL.

Esta REsL segue os princípios básicos da RSL apresentados por [12] [13] [14]. A técnica de *snowballing* (*backward* e *forward*) contribuiu para a cobertura do conjunto final de artigos selecionados [15]. *Snowballing* é uma técnica de busca de estudos sistemáticos da literatura, em que sua aplicação se dá por meio de uma busca na lista de referências de obras (*backward*) ou citações em máquinas de buscas (*forward*) [15].

## A. Planejamento da REsL

Nesta etapa, foram definidas duas questões de pesquisa (QP) para a realização desta revisão:

[QP1]: *Quais desafios são enfrentados pelos educadores no ensino de sistemas IoT utilizando jogos?* Nós definimos desafios como as dificuldades tecnológicas, problemas e obstáculos enfrentados pelos educadores no ensino da construção de sistemas IoT com uso de jogos.

[QP2]: *Quais são as oportunidades encontradas no ensino de sistemas IoT utilizando jogos?* As oportunidades indicam as possíveis lacunas identificadas a serem seguidas como trabalhos futuros na área de jogos e sistemas IoT.

As QPs foram organizadas no GQM - *Goal Question Metric* [16] para **analisar** o ensino de sistemas IoT usando jogos **com o propósito de** caracterizar **em relação aos** desafios e oportunidades **do ponto de vista dos** pesquisadores da engenharia de software **no contexto** do conhecimento disponível na literatura técnica.

Além disso, o PICO (População, Intervenção, Comparação e Resultado) [17] foi utilizado para organizar a *string* de busca. Diversos estudos secundários utilizam o PICO como forma de auxiliar na formulação da questão de pesquisa, onde a *string* de busca é definida. A Scopus[1] foi escolhida como máquina de busca para encontrar os artigos, uma vez que a mesma indexa outras máquinas de busca. Além disso, a técnica *snowballing* (mencionada anteriormente) foi incorporada ao processo de busca para mitigar as possíveis lacunas que poderia haver na aplicação da *string* de busca, e assim, ter uma maior cobertura na busca pelos artigos. Os artigos selecionados estavam entre os anos de 2015 e 2020. A string de busca está apresentada na Tabela 1. O protocolo de pesquisa completo está disponível em *https://bit.ly/3qNtZx1*.

TABELA 1. ORGANIZAÇÃO DA *STRING* DE BUSCA

| | **String de Busca** |
|---|---|
| P | "Internet of Things" OR IoT |
| I | Education* OR Teach* OR Learn* OR Training |
| C | Não aplicado |
| O | Game OR "Serious games" OR Gami* OR "Game-based Learning*" |

## B. Execução da REsL

A REsL foi executada de outubro a dezembro de 2020 por dois pesquisadores (um pesquisador com nível de mestre e outro com nível de doutor). Procurou-se artigos de 2015 a 2020, uma vez que o paradigma dos sistemas IoT teve grandes avanços nesse período. O procedimento de seleção foi realizado por um pesquisador que avaliou cada estudo e um segundo que verificou os procedimentos, assim, evitando viés do pesquisador na seleção dos trabalhos. Com base nos critérios de aceitação definidos no protocolo, o pesquisador selecionou os estudos da seguinte forma:

1. o pesquisador aceitou o artigo: o estudo é incluído; e.
2. o pesquisador ficou em dúvida sobre o artigo: o estudo é incluído; e
3. o pesquisador rejeitou o artigo: o estudo é excluído.

Os critérios de inclusão e exclusão para a seleção dos artigos são definidos na Tabela 2.

TABELA 2. CRITÉRIOS UTILIZADOS NA SELEÇÃO DOS ARTIGOS

| | **Critérios de inclusão** |
|---|---|
| 1 | O artigo deve estar no contexto de sistemas IoT |
| 2 | O artigo deve relatar um estudo primário ou secundário |
| 3 | O artigo deve relatar um estudo baseado em evidências fundamentado em métodos empíricos (por exemplo, entrevistas, pesquisas de opinião, estudos de caso, experimento formal, etc.) |
| 4 | O artigo deve fornecer dados para responder, pelo menos, uma das questões de pesquisa do REsL |
| 5 | O artigo deve ser escrito no idioma inglês |
| | **Critérios de exclusão** |
| 1 | Publicação duplicada/autoplágio |
| 2 | Anais e pôsteres |
| 3 | Artigos que não são revisados por pares |

A extração de dados visa obter informações dos artigos selecionados para responder às questões de pesquisa. Os campos do formulário de extração de dados são enumerados na Tabela 3. As informações extraídas foram divididas em duas categorias: Dados Demográficos e Questões de Pesquisa (QP1 e QP2).

TABELA 3. FORMULÁRIO DE EXTRAÇÃO

| *ID do artigo* | | |
|---|---|---|
| Título | Título do artigo | Dados demográficos |
| Abstract | Abstract do artigo | Dados demográficos |
| Fonte de Publicação | Conferência, *Journal* ou Simpósio em que o artigo foi publicado | Dados demográficos |
| Ano | Ano da publicação do artigo | Dados demográficos |
| Tipo de estudo | Estudo primário ou secundário identificado no artigo | Dados demográficos |
| Contexto | Tipo de evidência | Dados demográficos |
| Desafios | Descreve os desafios em aberto na prática de ensino de IoT usando jogos | QP 1 |
| Oportunidades | Descreve as oportunidades abertas no ensino de IoT com jogos | QP 2 |

---

[1]Scopus.com

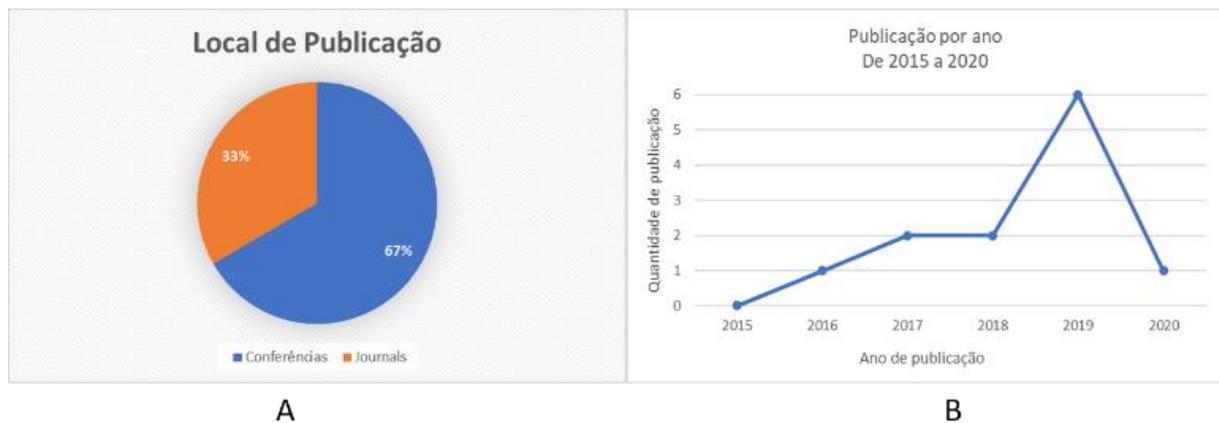

FIGURA 1. (A) MOSTRA O LOCAL DE ORIGEM DAS PUBLICAÇÕES DOS TRABALHOS. (B) APRESENTA OS ARTIGOS PUBLICADOS POR ANO

## IV. RESULTADOS E DISCUSSÕES

Os resultados deste trabalho foram analisados de acordo com as QPs. O objetivo foi identificar informações essenciais, padrões de ensino e similaridades entre os conceitos extraídos dos estudos primários. São discutidas, concisamente, as principais descobertas para cada questão de pesquisa a seguir. Os resultados da REsL mostram que existem poucos trabalhos relacionados ao uso de jogos para ensino de sistemas IoT. De acordo com os resultados obtidos pelos dados demográficos, a Figura 1 mostra os trabalhos publicados ao longo dos anos sobre o ensino de sistemas IoT apoiados por jogos. Além disso, é apresentado a fonte de publicação dos trabalhos selecionados.

A busca resultou 455 artigos, com 256 restantes após a remoção de duplicatas, pôsteres e procedimentos. Em seguida, um pesquisador fez a triagem dos artigos por Título, resultando em 81 artigos selecionados. Em seguida, foi aplicada a seleção baseada em resumo, na qual foram excluídos 53 artigos. Como resultado, 28 artigos foram incluídos após esta etapa. Por fim, após a aplicação da leitura na íntegra, nove artigos foram selecionados para compor o conjunto final de artigos, bem como a aplicação da técnica *Snowballing*. Após a aplicação do *snowballing*, três artigos foram adicionados, resultando 12 artigos para o conjunto final. A Tabela 4 mostra os artigos selecionados. O protocolo de pesquisa disponível detalha a etapa de seleção dos artigos.

TABELA 4. ARTIGOS SELECIONADOS

| ID do artigo | Título do artigo |
|---|---|
| P1 | Analysis on educating mechanical engineers through serious games using pervasive technologies |
| P2 | Concept for introducing the vision of industry 4.0 in a simulation game for non-IT students |
| P3 | Game-based learning for IoT: The tiles inventor toolkit |
| P4 | A Game-Based Learning System for Plant Monitoring Based on IoT Technology |
| P5 | A Serious Game for Competence Development in Internet of Things and Knowledge Sharing |
| P6 | An educational IoT lab kit and tools for energy awareness in European schools |
| P7 | ArViz: An IoT Teaching Tool for High School Students |
| P8 | Designing an IoT-focused, Multiplayer Serious Game for Industry 4.0 Innovation |
| P9 | Orgatronics: A physically interactive videogame for learning biology concepts using IoT technologies |
| P10 | Aligning Education for the Life Sciences Domain to Support Digitalization and Industry 4.0 |
| P11 | Using an interdisciplinary demonstration platform for teaching Industry 4.0 |
| P12 | A Method to Diagnose, Improve, and Evaluate Children's Learning Using Wearable Devices Such as Mobile Devices in the IoT Environment |

**SQP1: Quais desafios são enfrentados pelos professores no ensino de sistemas IoT utilizando jogos?**

Em relação aos desafios, alguns artigos [P3, P5 e P8] apresentam o uso de sistemas IoT com jogos para o ensino-aprendizagem. Os autores desenvolveram jogos sérios para a aplicação dos estudos. Alguns destaques como restrições de hardware e configuração de ambientes foram identificados pelos autores.

No trabalho apresentado em [P7], os autores utilizam a plataforma *ArViz* para ensinar programação a programadores novatos e inexperientes com o paradigma IoT. Os desafios encontrados eram relacionados à programação em módulos. Pelo fato da plataforma *ArViz* utilizar uma interface gráfica e baseada em blocos, o que não requer conhecimento profundo de programação a priori.

Os artigos [P6 e P12] apresentam um estudo de caso voltado para apoiar crianças e adolescentes a usarem sistemas IoT. No entanto, o foco não é o ensino de sistemas IoT, mas sim, como utilizá-los e montá-los com o objetivo de ensinar outro assunto. Os autores apresentam desafios em relação à associação das "coisas" e com os circuitos digitais.

Pode-se observar, há alguns desafios a serem abordados, principalmente no que diz respeito ao ensino de programação e restrições de hardware no contexto da IoT. Os trabalhos, em sua maioria, não tratam de desafios específicos para ensinar a construir tais sistemas. Eles somente utilizam o paradigma IoT, pelo fato dele ser multidisciplinar, para alcançar algum outro tipo de objetivo, como por exemplo, economia de energia, ou, até mesmo, em apoio ao ensino de outras áreas.

**SQP2: Quais são as oportunidades encontradas no ensino de sistemas IoT utilizando jogos?**

Em relação as oportunidades, temos que alguns autores [P4 e P9] utilizaram os sistemas IoT combinados com a aprendizagem baseada em jogos como apoio de ensino da área de botânica. Isso mostra um cenário totalmente inovador para o contexto da Indústria 4.0, apesar dos trabalhos não serem específicos para ensino de IoT.

Os artigos [P6, P7 e P12] descrevem o ensino de sistemas IoT para crianças e adolescentes em escolas de ensino fundamental e médio. Em geral, a ideia é introduzir sistemas IoT desde a infância para que as crianças e adolescentes se familiarizem com o novo paradigma.

Os trabalhos [P2, P10 e P11] relatam estratégias de ensino de sistemas IoT a estudantes de outras áreas, além da computação. O objetivo é fazer com que outras áreas tenham uma visão geral sobre as características e tendências da nova revolução industrial. O trabalho apresentado em [P1] usa uma abordagem baseada em jogos sérios para ensinar estudantes de engenharia mecânica sobre aplicações pervasivas. Apesar de aplicações pervasivas serem diferentes de sistemas IoT, ainda assim, ambos os sistemas compartilham características em comum devido a eles estarem dentro da Indústria 4.0.

Sistemas IoT, em sua essência, possuem características multidisciplinares apoiando outras áreas de estudos. Isso pode estar relacionado ao fato de tais sistemas estarem dentro do contexto da Indústria 4.0. No entanto, notamos poucos trabalhos com uso de jogos para apoiar o ensino de sistemas IoT em um contexto da área de computação. Esse fator pode estar associado com a novidade e características que esses sistemas trazem consigo. Além disso, mapear os possíveis cenários para construção de jogos específicos para IoT pode ser uma tarefa desafiadora, uma vez que eles são sistemas dinâmicos, que se adaptam ao ambiente de acordo com os dados coletados do ambiente em que estão imersos. A incorporação do ensino dos sistemas IoT nos ensinos primários no Brasil pode ser vantajoso, pois esses sistemas se tornarão cada vez mais comum no dia a dia. Então, fazer com que crianças e adolescentes lidem com eles durante as aulas seja uma ideia promissora.

V. AMEAÇAS À VALIDADE E LIMITAÇÕES DA PESQUISA

Este estudo teve algumas limitações: (i) poucos estudos específicos sobre o ensino de sistemas IoT com apoio de jogos. Assim, a análise foi limitada pelo número de artigos encontrados; (ii) embora tenhamos executado a REsL em um único motor de busca, nós complementamos com a técnica *snowballing*; (iii) a generalização dos resultados. Não podemos garantir que os dados coletados e interpretados sejam aplicáveis em todos os domínios contidos na prática de ensino de sistemas IoT.

Em relação a ameaça à validade externa, não podemos generalizar os desafios e oportunidades encontradas nos estudos para toda a população de ensino de sistemas IoT, uma vez que os estudos trataram apenas de domínios específicos, no caso, somente a aplicação de jogos para ensinar sistemas IoT e não outros meios de ensino, por exemplo, não utilizando jogos como apoio. Outra ameaça identificada neste trabalho foi o viés de interpretação dos pesquisadores, uma vez que este estudo se caracteriza como qualitativo.

Os estudos secundários devem, em sua essência, ter seus dados rastreáveis. Para mitigar esta ameaça, nossos resultados estão disponíveis no protocolo de pesquisa para eventuais consultas e replicações. Além disso, os resultados foram analisados e revisados por dois pesquisadores para reduzir ao máximo tal viés.

Em relação à confiabilidade, cada etapa deste estudo é descrita no protocolo de pesquisa. Este protocolo pretende promover sua replicação em estudos futuros. A *string* de busca também pode ser um fator de ameaça, pois uma *string* inadequada pode não ser capaz de obter os artigos necessários. Para diminuir esta ameaça, a *string* foi revisada por um grupo de pós-graduandos que cursaram uma disciplina específica em que este estudo foi desenvolvido.

VI. CONSIDERAÇÕES FINAIS E TRABALHOS FUTUROS

Este trabalho apresentou uma visão geral dos principais desafios e oportunidades identificados na literatura técnica sobre o uso de jogos no contexto de ensino de sistemas IoT nos últimos cinco anos. Foi realizada uma Revisão Estruturada da Literatura para encontrar e obter alguns desafios que educadores se deparam para ensinar a construir tais sistemas, bem como algumas lacunas direcionando a novos campos de pesquisas.

Os sistemas IoT surgiram como uma subcategoria da Indústria 4.0, tornando-se foco de diversos estudos na última década. Apontamos um conjunto de desafios diretamente associados à combinação de jogos e sistemas IoT, como a manipulação e montagem das coisas, ensino de programação e modularidade, configurações de ambientes, as limitações de recursos dos dispositivos, hardware e "coisas", dente outros.

Como oportunidades de pesquisa, pode-se destacar o ensino de sistemas IoT apoiado por jogos não limitado somente a alunos de graduação. Iniciativas na Europa e Ásia estão implementando o uso de sistemas IoT no ensino fundamental e médio. Outra oportunidade, é a aplicação de ensino de IoT com jogos em outras áreas além da computação, como a área de engenharia mecânica, botânica e administração, dente outros.

Em relação aos trabalhos futuros, podemos destacar: (i) Reexecução desta pesquisa, com o objetivo de evoluir o estado da arte na área de ensino de sistemas IoT em um contexto além da utilização de jogos. (ii) Expandir e aplicar os critérios de inclusão/exclusão em trabalhos escritos em português (como SBIE, RBIE, dentre outros). (iii) Realizar uma revisão da literatura cinza, olhando o estado da prática. E, por fim, (iv) pretende-se propor um guia para criação de jogos específicos para sistemas IoT, considerando os domínios e as características únicas que esses sistemas possuem.

REFERÊNCIAS

[1] R.C. Motta, K.M. de Oliveira, and G.H. Travassos, "On Challenges in Engineering IoT Software Systems". In: Anais do XXXII Simpósio Brasileiro de Engenharia de Software, 2018, pp. 42-51.

[2] Y. Liao, F. Deschamps, E.D.F.R., Loures, and L.F.P., Ramos, "Past, present and future of Industry 4.0 - a systematic literature review and

research agenda proposal". International Journal of Production Research, v.55, n. 12, 2017, pp. 3609–3629.

[3] S. Matalonga, D. Amalfitano, A. Doreste, A.R. Fasolino, and G.H. Travassos, "Alternatives for Testing of Context-Aware Contemporary Software Systems in industrial settings: Results from a Rapid review". 2021. arXiv preprint arXiv:2104.01343.

[4] F. Zambonelli, "Key abstractions for IoT-oriented software engineering". IEEE Software, v. 34, n. 1, 2017, pp. 38-45.

[5] E.O. Navarro, A. Baker, and A.V.D. Hoek, "Teaching software engineering using simulation games." In: ICSIE'04: Proceedings of the 2004 International Conference on Simulation in Education. San Diego, CA, s.n. 2004.

[6] L. Atzori, A. Iera, and G. Morabito, The Internet of Things: A survey. Computer Networks, v. 54, n. 15, 2010, pp. 2787–2805

[7] R.C. Motta, V. Silva, and G.H. Travassos, "Towards a more in-depth understanding of the IoT Paradigm and its challenges." Journal of Software Engineering Research and Development, v. 7, 2019, pp. 3-16.

[8] G. Xexeo, A. Carmo, A. Acioli, B. Taucei, C. Dipolitto, E. Mangeli, and R. Monclair, O que são jogos. LUDES. Rio de Janeiro, 1, pp. 1-30.

[9] D. Castro, D. Costa, and C. Werner, "Systematic mapping on the use of games for software engineering education". In: 23º Conferência Ibero-americana de Engenharia de Software (CIbSE), Curitiba, 2020, pp. 1-14.

[10] E. Adams, and A. Rollings, Fundamentals of game design (game design and development series). Prentice-Hall, Inc, 2006.

[11] K.M. Kapp, The gamification of learning and instruction: game-based methods and strategies for training and education. John Wiley & Sons, 2012.

[12] J. Biolchini, P.G. Mian, A.C.C. Natali, and G.H. TRAVASSOS, "Systematic review in software engineering". System Engineering and Computer Science Department COPPE/UFRJ, Technical Report ES, 679(05), 45.

[13] K. Petersen, R, Feldt, S. Mujtaba, and M. MATTSSON, "Systematic mapping studies in software engineering." In 12th International Conference on Evaluation and Assessment in Software Engineering (EASE), 2008, pp. 1-10..

[14] B. Kitchenham, and S. Charters, "Guidelines for performing systematic literature reviews in software engineering." Relatório Técnico Evidence-Based Software Engineering (EBSE), v. 2.3. 2007.

[15] C. Wohlin, "Guidelines for snowballing in systematic literature studies and a replication in software engineering." In: Proceedings of the 18th international conference on evaluation and assessment in software engineering. 2014, pp. 1-10. C.

[16] V.R. Basili, " Software modeling and measurement: The Goal/Question/Metric paradigm", 1992.

[17] M. Pai, et al. "Systematic reviews and meta-analyses: an illustrated, step-by-step guide." The National medical journal of India, v. 17, n. 2, 2004, pp. 86-95.

[18] J. Henry, S. Tang, M. Hanneghan, and C. Carter, "A framework for the integration of serious games and the internet of things (IoT)". In IEEE 6th International Conference on Serious Games and Applications for Health (SeGAH), 2018, pp. 1–8. IEEE.

[19] M. Kassab, J. DeFranco, and P. Laplante, "A systematic literature review on Internet of things in education: Benefits and challenges". Journal of Computer Assisted Learning, 36(2), pp. 115-127, 2020.

[20] K. Ashton, "That 'Internet of Things'" RFiD J., 2009

APÊNDICE. REFERÊNCIAS DOS ARTIGOS SELECIONADOS

P1. J.M. Baalsrud Hauge, T. Lim, M. Kalverkamp, F. Haase, and F. Bellotti, "Analysis on educating mechanical engineers through serious games using pervasive technologies". 2016, In International Design Engineering Technical Conferences and Computers and Information in Engineering Conference

P2. M. Zarte, and A. Pechmann, "Concept for introducing the vision of industry 4.0 in a simulation game for non-IT students". IEEE 15th International Conference on Industrial Informatics (INDIN). IEEE, pp. 512-517, 2017.

P3. A. Mavroudi, M. Divitini, S. Mora, and F. Gianni, F, "Game-based learning for IoT: The tiles inventor toolkit". In Interactive Mobile Communication, Technologies and Learning, pp. 294-305, 2017.

P4. P. Tangworakitthaworn, V. Tengchaisri, K. Rungsuptaweekoon, and T.A. Samakit, "Game-Based Learning System for Plant Monitoring Based on IoT Technology". In 2018 15th International Joint Conference on Computer Science and Software Engineering (JCSSE), pp. 1-5, 2018.

P5. U. Nima, R. Wangdi, J.B.A. Hauge, "Serious Game for Competence Development in Internet of Things and Knowledge Sharing". In IEEE International Conference on Industrial Engineering and Engineering Management (IEEM), pp. 1786-1790, 2018.

P6. G. Mylonas, D. Amaxilatis, L. Pocero, I. Markelis, J. Hofstaetter, and P. Koulouris, "An educational IoT lab kit and tools for energy awareness in European schools". International Journal of Child-Computer Interaction, 20, pp. 43-53, 2019.

P7. K. Chochiang, K. Chaowanawatee, K. Silanon, and T. Kliangsuwan, "ArViz: An IoT Teaching Tool for High School Students". In 23rd International Computer Science and Engineering Conference (ICSEC), pp. 87-91, 2019.

P8. M. Oliveri, J.B. Hauge, F. Bellotti, R. Berta, and A. De Gloria, "Designing an IoT-focused, Multiplayer Serious Game for Industry 4.0 Innovation". In IEEE International Conference on Engineering, Technology and Innovation (ICE/ITMC), pp. 1-9, 2019.

P9. D. Santiago, G. Chicangana, O. Santiago, and L. Erazo, "Orgatronics: A physically interactive videogame for learning biology concepts using IoT technologies". In: International Conference on Virtual Reality and Visualization (ICVRV). IEEE, pp. 138-141, 2019.

P10. C. Catal, and B. Tekinerdogan, "Aligning Education for the Life Sciences Domain to Support Digitalization and Industry 4.0". Procedia computer science, 158, pp. 99-106, 2019.

P11. J. Wermann, A.W. Colombo, A. Pechmann, and M. Zarte, "Using an interdisciplinary demonstration platform for teaching Industry 4.0". Procedia Manufacturing, v. 31, pp. 302-308, 2019.

P12. M. Moradi, and K. Rahsepar Fard, "A Method to Diagnose, Improve, and Evaluate Children's Learning Using Wearable Devices Such as Mobile Devices in the IoT Environment". Mobile Information Systems, v. 2020, 2020.